\documentclass{icrc}

\usepackage{times}

\usepackage{graphicx} % when using Latex and dvips

\begin{document}

\title{Very high-energy $\gamma$-ray observations of
the Crab nebula with the GRAAL experiment}

\author[1]{F.Arqueros}
\affil[1]{Facultad de Ciencias F\'{\i}sicas, Universidad Complutense,
         E-28040 Madrid, Spain}
\author[2]{J.Ballestr\'{\i}n}
\affil[2]{CIEMAT-Departamento Energ\'{\i}as Renovables, \\Plataforma Solar
de Almer\'{\i}a, E-04080 Almer\'{\i}a, Spain}
\author[3]{M.Berenguel}
\affil[3]{Departamento de Lenguajes y Computaci\'on, \\
Universidad de Almer\'{\i}a, 04120 Almer\'{\i}a, Spain}
\author[1]{D.M.Borque}
\author[4]{E.F.Camacho}
\affil[4]{Escuela Superior de Ingenieros, Universidad de Sevilla,
E-41012 Sevilla, Spain}
\author[5]{M.D\'{\i}az}
\affil[5]{Max-Planck-Institut f\"ur Physik, 80805 M\"unchen, Germany}
\author[1]{R.Enr\'{\i}quez}
\author[5]{H.-J.Gebauer}
\author[5]{R.Plaga}

% \titleheight{11cm} % uncomment and adjust in case your title block

                     % does not fit into the default and minimum 7.5 cm

\maketitle
\begin{abstract}
The ``Gamma Ray Astronomy at ALmer\'{\i}a'' (GRAAL) experiment
uses 63 heliostat-mirrors with a total mirror area
of  $\approx$ 2500 m$^2$
from the CESA-1 field to collect
Cherenkov light from airshowers.
The detector is located in a central solar tower and 
detects photon-induced showers
with an energy threshold of 250 $\pm$ 110 GeV and an 
asymptotic effective 
detection area
of about 15000 m$^2$.
\\
Data sets taken
in the period September 1999 - September 2000 in the direction of
the Crab pulsar were analysed
for high energy $\gamma$-ray emission.
Evidence for $\gamma$-ray flux from the Crab pulsar with an integral flux of 
2.2 $\pm$ 0.4 (stat) $^{+1.9}_{-1.5}$ (syst) $\times$ 
10$^{-9}$ cm$^{-2}$ s$^{-1}$ above threshold
and a significance of 4.5 $\sigma$
in a total (usable) observing time of 7 hours and 10 minutes on source
was found. No evidence for emission from the other sources was seen.
\\
The effect of field-of-view restricted to the central part
of a detected airshower on the lateral distribution and timing
properties of Cherenkov light and their effect on an efficient $\gamma$-hadron
separation are discussed. 
\end{abstract}

\section{Introduction}\label{sec_intro}
Measuring atmospheric Cherenkov radiation 
is presently the most effective way to  detect
cosmic $\gamma$-rays with primary energies between about 100 GeV and 1 TeV (\cite{frank}). 
In order to reach low energy thresholds with techniques  
based on Cherenkov light, large mirror collection areas are needed.  
GRAAL is an experiment that employs  
the large mirror area of an existing tower
solar-power plant for this purpose.
\section{The GRAAL detector}
\label{detector}

\begin{figure}[t]
\includegraphics[width=8.3cm]{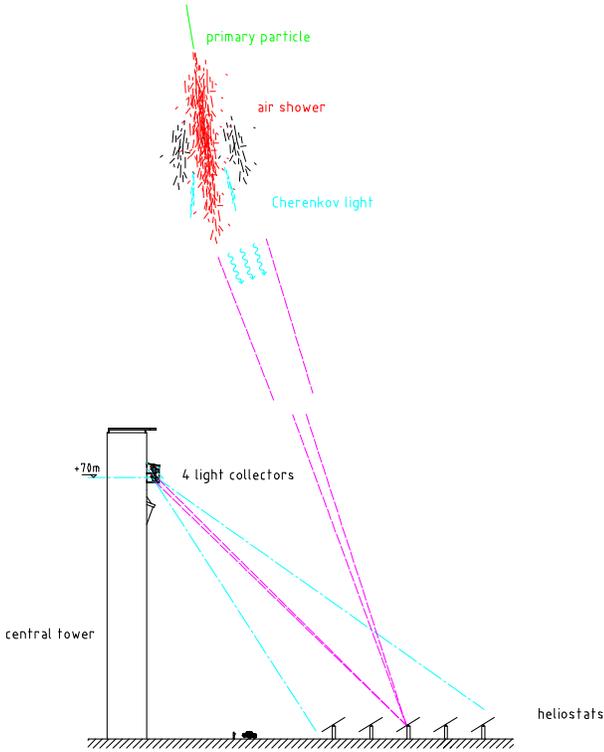}
\caption{\it
Scheme of the experiment seen from the side, north is to the right.
The Cherenkov light of a schematic airshower 
(not to scale with respect to the field)
is concentrated by the heliostats of the CESA-1 field
to a focus at the central tower.
A dedicated platform mounted at the outside of the tower at the 70 m level  
houses four Winston  
cones which receive light form 13 - 18 heliostats in the field. 
The data-acquisition electronics is located inside the tower.}
\label{tower}
\end{figure}

CESA-1 is a heliostat field comprising of 225 
steerable mirrors to the north
of a central tower located within the 
``Plataforma Solar de Almer\'{\i}a''(PSA), a solar thermal-energy 
research centre located in the desert of Tabernas 
(37$^{\circ}$.095 N, 2$^{\circ}$.360 W) at a
height a.s.l. of 505 m. 
The 63 heliostats used for GRAAL
have a mirror area of 39.7 m$^2$ each and have a roughly spherical shape.
\\
The heliostats focus the Cherenkov light of airshowers from
the direction of potential gamma-ray sources to 
software adjustable ``aiming points'' in the central tower 
(see fig.\ref{tower}). 
Cherenkov light from four groups of heliostats  
(with 13,14,18,18 members, respectively)   
is directed onto four single non-imaging ``cone concentrators'' (truncated
Winston cones with an opening angle of 10$^{\circ}$)
each containing a single large-area PMT.
They are housed in a special enclosure that is fastened to the
outside of the central tower at a height of 70 m.
The incoming light from an air shower consists of a train of   
pulses from the different heliostats, 
usually fully separated  
by pathlength differences. The arrival time and amplitude
of each heliostat can thus be determined with a flash-ADC in
a sequential mode.
\\
Compared to the three other heliostat field experiments
(``CELESTE''(\cite{celestecrab}),``STACEE''(\cite{staceetech}) and
``Solar 2'' (\cite{solar2})) which focus the light of single
heliostats onto related single PMT's, the night sky background (NSB) per
channel is about a factor 10 higher in GRAAL (8-10 p.e./ns). This leads to a
higher energy threshold.
The advantage of the non-imaging approach is its greater simplicity
leading to cost savings by about a factor 5-10 in hardware costs.
The presence of only  four data-acquisition 
channels makes automatization and
remote control more  feasible, leading to comparable
savings in operation costs. 
In the non-imaging approach it is impossible to avoid a temporal
overlap of the signal from certain heliostats depending on the
pointing direction. This reduces the number of times/amplitudes
usable in the reconstruction by about 20 $\%$. 
On the positive side, calibration is easier when signals from several
heliostats are measured in the same PMT.
\\
We register all four pulse trains  
in only one Digital Oscilloscope   
with a bandwidth of 1 GHz and a time bin of 500 psec.  
This ensures that the FWHM of individual pulses of about 3.6 nsec 
is negligibly increased by electronics effects.
The time and amplitude calibration 
of our setup is performed using blue LEDs with a calibrator  
module that is fastened at the window of the Winston cones. 
\\
All operations (like opening of the door, high-voltage 
control etc.) at the central receiver and the tracking 
of the heliostat field are under remote control via the internet. Under conditions
that indicate some malfunction, a physicist on shift is phoned by the PC
and can check all parameters and images of web cameras, remotely.
For the operation of the heliostat field and emergencies only the regular
night-operator of the PSA is on-site in all observation nights. 
GRAAL is taking data continously since August 1999 (since March 2000 in the
final hardware configuration). Here we present data for the observation year
1999/2000 of the Crab nebula.
\section{Event reconstruction}
\label{evreco}
\begin{figure}[t]
\includegraphics[width=8.3cm]{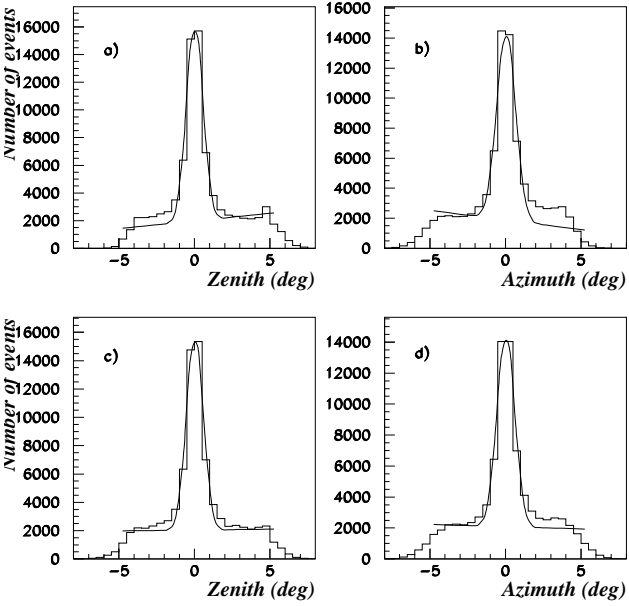}
\caption{Projections of the number of showers as a function
of shower directions as reconstructed from the timing data. 
Shown is deviation of the reconstructed direction 
from the pointing direction 
on the elevation-axis (panels a. and c.) and azimuth-axis 
(panels b. and d.). 
The origin corresponds to the pointing
direction as determined by the orientation of the heliostats.
The data sample comprises
of 32 hours of ON-source time on the Crab pulsar (panels a. and b.) 
and an equal amount of
OFF-source time (panels c. and d.) taken under variable 
weather conditions in the season
1999/2000. The ``Gaussian plus linear function'' fit 
is performed to each subsample.
It is seen that the Gaussian - corresponding to successfully 
direction reconstructed events -
is always centred within $<$ 0.05$^{\circ}$.}
\label{angproj}
\end{figure}

The different arrival times of the signals from the individual heliostats were
used to reconstruct the timing shower front of the individual events.
The expected arrival times for all heliostats in each of the
four cones were calculated and stored in a ``library''
for a 5 $\times$ 5 degree grid
centered to a direction about 1 degree offset from the current
pointing direction of the heliostats. The offset was chosen to
avoid a bias towards ``correct pointing''.
This calculation was performed assuming a point-like
shower-maximum at a penetration depth of 230 g/cm$^2$
(the mean penetration of showers induced by a photon of 100 GeV)
in the pointing direction. A spherical timing-front was assumed
to be emitted by this maximum. The shower core was fixed at the
geometrical centre of the field as defined by the used heliostats (it was
verified that this assumption introduces no bias). 
\\ 
The measured arrival
times were compared to the ``library''. We define the time difference
TIMEDIFF
%\begin{equation}
\begin{eqnarray}
\rm{TIMEDIFF=(measured\: arrival\: time)- } 
\nonumber \\ \rm{ (nearest\: expected\: time\: from\: the\: library)}
\label{timediff}
\end{eqnarray}
%\end{equation}
The direction yielding the smallest $\chi_t^2$
with \begin{equation}
\chi{_t}^2 =\sum_i ({\rm TIMEDIFF}_i)^2 
\label{chi2}
\end{equation} was chosen as the final 
reconstructed direction of the shower.
\\
Fig.\ref{angproj} shows projections of reconstructed directions
in zenith and azimuth angle both for ON and OFF source directions
for a large data sample. The origin corresponds to the
pointing direction determined by the heliostat tracking.
The directions of events in the ``smooth
background'' extending to large off-axis angles were found to be systematically
misreconstructed. This effect was used in the later analysis to normalise ON
and OFF rates.
\\
If the ``misreconstructed'' directions are excluded, the
angular resolution $\sigma_{63}$ (the opening angle within which
63 $\%$ of all events are contained) is 0.7$^{\circ}$.

\section{Data reduction}
\label{datared}
From a total measuring time of 32 hours, only nights in which all four detector channels
and the heliostats in the field were functioning normally according to the 
recorded monitor files 
were chosen for further analysis.
Furthermore, only data taken in ``good nights'', i.e., with no clouds, low
humidity, no dust, remained after our cuts.
These ``meteorological cuts'' are severe under the weather conditions
at the PSA. In the data sample on Crab in February/March 2000 
only 22 $\%$ of all data 
taken on the Crab pulsar passed all cuts.
\begin{table*}[t]
\footnotesize
\caption{\it Currents (mean of 4 Cones), single trigger rate of charge 
integrating channel (mean of Cone 1+2),
number of hardware-triggered events (``raw events''), decadic logarithm
mean net-charge of all events in sample (``mean q''),
number of events after angular reconstruction and software trigger (``rec. events'')
and normalised number of events in central
angular region (within 0.7 degrees of pointing direction) (``centr. events'')
for the samples with pointing towards
the Crab pulsar (``ON'') and on a sky position (``OFF'')
with a right ascension 2.625 degrees
smaller than in the ON direction.
The total data-taking time ON was 430 minutes with an equal amount of OFF time.}
\label{crabresult}

\vspace{0.1cm}
\begin{center}
\begin{tabular}{lcccccc}
  &  current [$\mu$A] & q-rate[kHz] &  mean q & raw events &  rec. events &  centr. events   \\
\hline
\hline
ON  & 19.0 $\pm$ 0.4 & 1.35 & 2.883 $\pm$ 0.004 & 68702   & 33384  & 9415    \\ 
OFF & 19.3 $\pm$ 0.3 & 1.49 & 2.876 $\pm$ 0.004 & 75198   & 33056  & 8678    \\ 
\hline
EXCESS & -0.3 & -0.14 & 0.007 $\pm$ 0.006 & -6496 & 328 $\pm$ 258 & 737 $\pm$ 165 \\ 
\hline
\end{tabular}
\end{center}
\normalsize
\end{table*}
\\
The fundamental problem of all Cherenkov experiments - especially 
for those attempting to detect an excess due to gamma-rays in the
total rate - is the fact that the night-sky background ON- resp. OFF-source
differs in general. This can influence the counting rate and 
analysis efficiency in various ways.
Whereas random events can be removed (see section \ref{raw}), the change of the
effective trigger threshold due to different noise levels in ON and OFF is a
more serious problem. Here, it was solved by applying a variable offline threshold,
calculated from the measured noise level in the related traces.
\\
One can attempt to find an ON-source excess in the total number of events (see
section \ref{raw} below). A more sensitive method is to look for an excess in
the central angular region. The normalised excess EXCESS$_{\rm{n}}$ was calculated according to the following equation:
\begin{equation}
\rm{EXCESS_n = ON_{in} - OFF_{in} \left(ON \over OFF\right)_{out}} 
\label{exc}
\end{equation}
Here $\rm{(ON,OFF)_{in}}$ stands for the number of events within 0.7$^{\circ}$
from the source, resp. off-source  
direction, $\rm{(ON,OFF)_{out}}$ stands for the number of 
events with directions
deviating more than 2$^{\circ}$ from the source direction.
\section{Results}
\label{results}
\begin{figure}[t]
\includegraphics[width=8.3cm]{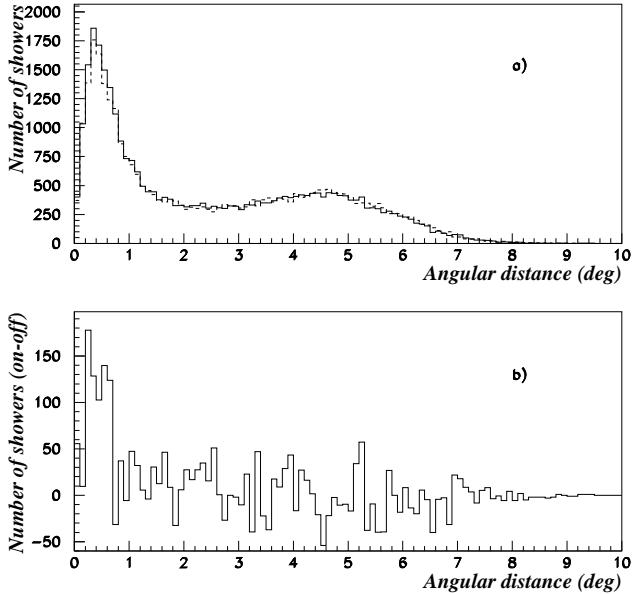}
\caption{\it
The upper plot (a.) shows the
number of events as a function of angular distance of 
reconstructed direction from source direction for ON-source events
(full line) and OFF-source events (dashed line).
No normalisation
of any kind was applied to this plot.
The lower plot (b.) shows the difference ON - OFF, normalised
to the number of events in the outer angular region, according to
eq. (3).}
\label{evangcrab}
\end{figure}

Several parameters of the data set taken on Crab pulsar are
presented in table \ref{crabresult}.
Fig. \ref{evangcrab} shows the number of events as function
of angular distance from the source direction, both for ON- and OFF-source
direction and the normalised difference ON-OFF. An 
excess of events in the angular region expected from Monte Carlo (MC) simulations
(fig.\ref{evangcrab}) is seen, we find EXCESS$_n$ = 737 $\pm$ 165 
calculated according to eqn (\ref{exc}). The error is statistical.
This corresponds to a 4.5 $\sigma$ excess and a mean excess rate
EXCESS$_{\rm nr}$ = 1.7/min.
%Fig. \ref{evang3c454} presents a data set of comparable size and quality
%with pointing towards the extragalactic potential gamma-source 3C 454.3.
An integral flux $\phi_{int}$ is calculated from this excess according to:
\begin{equation}
{\rm \phi_{int} = (EXCESS_{nr}/r_{\gamma}) 
(r_p/r_{\rm obs}) t_c \phi_{whipple}}
\label{flux}
\end{equation}
Here $\phi_{\rm whipple}$=
$\int_{\rm E_{\rm thresh}}^{\infty}$ 
3.3 $\times$ 10$^{-7}$ E$^{-2.4}$ m$^{-2}$s$^{-1}$ TeV$^{-1}$ dE
is the integral gamma-ray flux from the Crab above a threshold energy 
E$_{\rm thresh}$
as observed by the Whipple collaboration (\cite{whipple}). r$_{\gamma}$ is the 
gamma-ray rate expected in GRAAL from the MC simulated 
effective area for gammas 
based on this flux (0.011 Hz). Note that the absolute Whipple flux cancels in
eq. (\ref{flux}), and we only adopt the spectral index from ref. (\cite{whipple}).
r$_p$ is the proton rate expected in GRAAL
on the basis of the known absolute differential flux of cosmic-ray protons $\phi_{ref}$ 
and the effective 
area for protons (4.0 Hz). 
%\begin{figure}[ht]
%\epsfig{
%file=evangcrab.eps,width=9cm,height=7cm,clip=,angle=0}
%\caption{\it  
%\end{figure}
r$_{obs}$ is
the observed cosmic-ray rate in the final reconstructed sample, corrected
for dead time (1.6 Hz)   
.
The factor (r$_{\rm obs}$/r$_p$) is an
empirical correction for the fact that
our MC calculated 
proton effective area  
predicts a somewhat higher proton rate
than observed. 
t$_c$ is a correction factor for the fact that some photons are
expected in the ``outer angular region'' and was determined as
2.2 from MC data.
The final integral flux above threshold
assuming a differential spectral source index of -2.4 is:
\\
$\phi_{int}$ = 2.2 $\pm$ 0.4 (stat) $^{+1.9}_{-1.5}$ (syst) $\times$ 
10$^{-9}$ cm$^{-2}$ s$^{-1}$ above threshold
%\verb
\balance
\section{Excess in total rate}
\label{raw}
If the detected excess (discussed in section \ref{results}) is real, one can
estimate that there
should be an excess of $\approx$ 2270 events within our measuring
time. On the other hand, extrapolating the Whipple flux for
Crab nebula (\cite{whipple}) at our energy threshold, only 355 excess events
are expected.
\\
Due to
our trigger setup it can happen that the NSB triggers events if the NSB is
high. The rate of accidental events can be calculated from the single rates and subtracted from the
total rate. 
Other corrections are related to the dead time of the setup.
Table \ref{accident} right column shows the results of a careful correction for these effects for the data of the
analyzed sample of Crab (see section \ref{datared}). In the last column all the effects have
been corrected and the total time of measurement is 430 min in ON position and
the same time for OFF position. There is an excess in the OFF position
of 7234 events in the hardware-triggered events. After subtraction of
accidental events and corrections for dead time, the excess in the OFF position
is only 443 events, which is within the statistical fluctuations. For
orientation, a difference in the energy threshold of cosmic-ray protons
between ON and OFF of only
5GeV at an energy threshold of 2TeV already produces a difference of 550 events
for the same time of measurement and using the known cosmic-ray proton flux and a constant effective area of 8000 m$^2$.
\begin{table}[t]
\footnotesize
\caption{\it Number of hardware-triggered events (``total events''), number of
  events once subtracted 
  the accidental events and corrected for the dead time (``total corrected events'') }
\label{accident}
\vspace{0.1cm}
\begin{center}
\begin{tabular}{lcccccc}
  &  Total events & Tot. correct. evs \\
\hline
\hline
ON  & 79194   & 58107 \\ 
OFF & 86428   & 58550 \\ 
\hline
EXCESS & -7234 $\pm$ 575 &   -443 $\pm$ 483  \\ 
\hline
\end{tabular}
\end{center}
\normalsize
\end{table}   
\\
In an alternative approach the software analysis rejects accidental
events due to the low number of peaks and uncorrelated times of the peaks of
such events. Less than 0.6$\%$ of the accidental events pass the analysis for
a very similar NSB to the one of Crab sample. In our
analysis, a higher NSB rejects more accidental events. As seen in the
column 5 of table \ref{crabresult} also here there is no significant excess in the total rate. 
This lack of an excess in the total rate seems to cast a doubt on the reality
of the signal discussed in section \ref{results}.
%This flux was derived in a way that links the absolute light calibration
%directly to the rate of hadronic cosmic rays.
\section{Conclusion}
\label{concl}

The results of the present measurements do not prove that the use
of an heliostat array in gamma-ray astronomy is a feasible alternative
to the use of dedicated Cherenkov telescopes. 
\\
The principle drawbacks of this approach were found
to be the restricted field of view and the weather conditions 
at the relatively low elevation of the heliostat field.
The field-of view restriction leads to a very
similar time structure of the shower front in proton and gamma induced
showers and biases the direction reconstruction based in timing
towards to the pointing direction. Both effects together prevent any efficient
separation of proton and gamma induced showers. 
This makes a flux determination independent of total rates difficult
(though not impossible) and severely limits the sensitivity of the experiment.
The fraction of time (total duty cycle)
with weather and moon-light conditions sufficient
for the detection of gamma radiation was about 3-4 $\%$ at the PSA, about
a factor of 5 lower than at astronomical sites. 
Both drawbacks seem to be unavoidable for the heliostat-field based approach also in the future.

\begin{acknowledgements}
The GRAAL project is supported by funds from the DFG, CICYT and the 
IHP ``Access
to large scale facilities'' program of the EU.
We thank the PSA - in particular A.Valverde - for excellent working 
conditions at the CESA-1 heliostat field.
\end{acknowledgements}

\end{document}